\documentclass[aps,prl,showpacs,twocolumn,floats,psfig]{revtex4}
\usepackage{amssymb}
\usepackage{amsbsy}
\usepackage{amsmath}
\usepackage{epsfig}
\newcommand\bea{\begin{eqnarray}}
\newcommand\eea{\end{eqnarray}}
\newcommand\beq{\begin{equation}}
\newcommand\eeq{\end{equation}}
\newcommand\bib{\bibitem}

\newcommand{\non}{\nonumber}
\newcommand{\al}{\alpha}
\newcommand{\be}{\beta}
\newcommand{\de}{\delta}
\newcommand{\si}{\sigma}

\newcommand{\da}{\dagger}
\newcommand{\la}{\langle}
\newcommand{\ra}{\rangle}
\newcommand{\vk}{\vec k}
\newcommand{\vn}{\vec n}
\newcommand{\vcr}{\vec r}

\begin{document}

\title{Exact results for quench dynamics and defect production in a
two-dimensional model}

\author{K. Sengupta$^1$, Diptiman Sen$^2$ and Shreyoshi Mondal$^1$}
\affiliation{$^1$ TCMP division, Saha Institute of Nuclear Physics,
1/AF Bidhannagar, Kolkata 700 064, India \\ $^2$ Center for High Energy
Physics, Indian Institute of Science, Bangalore, 560 012, India}

\date{\today}

\begin{abstract}
We show that for a $d$-dimensional model in which a quench with a rate 
$\tau^{-1}$ takes the system across a $d-m$ dimensional critical surface, 
the defect density scales as $n \sim 1/\tau^{m\nu/(z\nu +1)}$, where 
$\nu$ and $z$ are the correlation length and dynamical critical exponents 
characterizing the critical surface. We explicitly demonstrate that the 
Kitaev model provides an example of such a scaling with $d=2$ and 
$m=\nu=z=1$. We also provide the first example of an exact calculation of 
some multispin correlation functions for a two-dimensional model which can 
be used to determine the correlation between the defects. We suggest 
possible experiments to test our theory.
\end{abstract}


\maketitle

Quantum phase transitions have been studied extensively for several years
\cite{sachdev1}. Such a transition is accompanied by diverging length and 
time scales \cite{sachdev1} leading to the absence of adiabaticity close to 
the quantum critical point. Thus the system fails to follow its instantaneous
ground state when some parameter in its Hamiltonian is varied in time at a 
finite rate $1/\tau$ which takes the system across the critical point. Since 
the final state of the system does not conform to the ground state of
its final Hamiltonian, defects are produced \cite{kz1,bd1}. The defect
density $n$ depends on the quench time $\tau$ as $n \sim 1/
\tau^{d\nu/(\nu z+1)}$, where $\nu$ and $z$ are the correlation length and
dynamical critical exponents at the critical point \cite{anatoly1}. A
theoretical study of such a quench dynamics requires a knowledge of the
excited states of the system. Such studies have therefore been mostly
restricted to phase transitions in exactly solvable models in one or infinite
dimensions \cite{ks1,das1,levitov1,bd2}. Experimental studies of
defect production due to quenching of the magnetic field in a two-dimensional
(2D) spin-1 Bose condensate have been undertaken \cite{sk1}. However,
exact studies of quench dynamics have not been carried out so far for
2D spin models.

In this Letter, we carry out such a study for the 2D Kitaev model 
\cite{kitaev1}.
We show that when the quench takes the system across a critical (gapless)
line, the density of defects scales as $1/\sqrt{\tau}$. The Kitaev model
has $d=2$ and $\nu = z =1$; hence, the above scaling is in contrast to the 
$n \sim 1/\tau$ behavior expected when the system passes through a critical 
point \cite{anatoly1}. In this context we provide a general discussion of 
the scaling of the defect density for a general $d$-dimensional model with 
arbitrary $\nu$ and $z$; we show that when the quench takes such a system
through a $d-m$ dimensional gapless (critical) surface, the defect density 
scales as $n \sim 1/\tau^{m\nu/(z\nu+1)}$. This result is a generalization
of the result of Ref. \cite{anatoly1} and hence constitutes a 
significant extension of our current understanding of defect production due 
to a quench. We also compute exactly some multispin correlation functions for 
our model, and use them to study the variation of defect correlations with 
the quench rate and the model parameters. Such an exact analysis 
of defect correlations has not been carried out so far for 2D systems.

\begin{figure}
\rotatebox{0}{\includegraphics*[width=\linewidth]{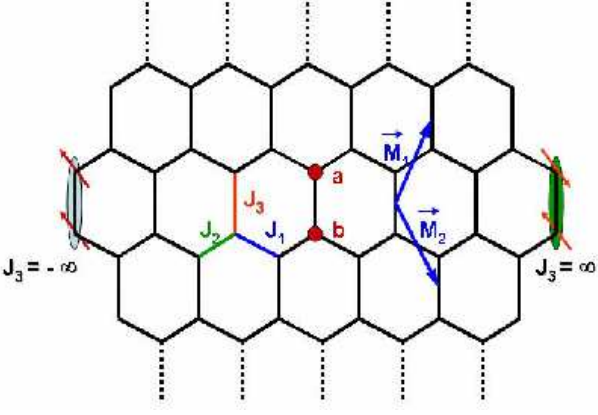}}
\caption{Schematic representation of the Kitaev model on a honeycomb lattice
showings the bonds $J_1$, $J_2$ and $J_3$. Schematic pictures of the ground
states, which correspond to pairs of spins on vertical bonds locked parallel
(antiparallel) to each other in the limit of large negative (positive) $J_3$,
are shown at one bond on the left (right) edge respectively. ${\vec M}_1$ and
${\vec M}_2$ are spanning vectors of the lattice, and $a$ and $b$ represent
inequivalent sites.} \label{fig1}
\end{figure}

The Kitaev model is a spin-1/2 model on a 2D honeycomb lattice with the 
Hamiltonian \cite{kitaev1}
\beq H = \sum_{j+l={\rm even}} ( J_1 \si_{j,l}^x \si_{j+1,l}^x + J_2
\si_{j-1,l}^y \si_{j,l}^y + J_3 \si_{j,l}^z \si_{j,l+1}^z ), \label{ham1} \eeq
where $j$ and $l$ denote the column and row indices of the lattice. This
model, sketched in Fig. \ref{fig1}, is known to have several interesting
features \cite{feng1,baskaran1,chen1,lee1}. It is a rare example of a 2D
model which can be exactly solved \cite{kitaev1,feng1,chen1}. It supports 
a gapless phase for $|J_1-J_2| \le J_3 \le J_1+J_2$ \cite{kitaev1} which
is possibly connected to a spin liquid state and demonstrates fermion 
fractionalization at all energy scales \cite{baskaran1}. In certain
parameter regimes, the ground state exhibits topological order and the 
low-energy excitations carry Abelian and non-Abelian fractional statistics; 
these excitations can be viewed as robust qubits in a quantum computer 
\cite{kitaev2}. There have been proposals for experimentally realizing
this model in systems of ultracold atoms and molecules trapped in optical
lattices \cite{expt_real}; such systems are known to provide easy access
to the study of non-equilibrium dynamics of the underlying model. However, 
in spite of several studies of the phases and low-lying excitations of 
the Kitaev model, its non-equilibrium dynamics has not been studied so far.
We will study what happens in this model when $J_3$ is varied from 
$-\infty$ to $\infty$ at a rate $1/\tau$, keeping $J_1$ and $J_2$ fixed.

One of the main properties of the Kitaev model which makes it theoretically
attractive is that, even in 2D, it can be mapped onto a non-interacting
fermionic model by a suitable Jordan-Wigner transformation \cite{feng1,chen1},
\beq H_F = i \sum_{\vn} [J_1 b_{\vn} a_{{\vn} - {\vec M}_1} + J_2 b_{\vn}
a_{{\vn} + {\vec M}_2} +J_3 D_{\vn} b_{\vn} a_{\vn}], \label{ham2} \eeq
where $a_{\vn}$ and $b_{\vn}$ are Majorana fermions sitting at the top and
bottom sites respectively of a bond labeled $\vn$, $\vn = {\sqrt 3} {\hat i} ~
n_1 + (\frac{\sqrt 3}{2} {\hat i} + \frac{3}{2} {\hat j} ) ~n_2$ denote the
midpoints of the vertical bonds shown in Fig. \ref{fig1}, and $n_1, n_2$ run 
over all integers. The vectors $\vn$ form a triangular lattice. The $x, y$ 
coordinates of the triangular lattice sites are given by $x= \sqrt{3}(n_1 +
n_2/2)$ and $y=3n_2/2$. [We will refer to the sites of the honeycomb lattice
either as $(j,l)$ as in Eq. (\ref{ham1}), or as $(a,\vn)$ and $(b,\vn)$
as in Eq. (\ref{ham2}); $j+l$ is even (odd) for $a$ ($b$) sites
respectively.] The vectors ${\vec M}_1 = \frac{\sqrt 3}{2} {\hat i} +
\frac{3}{2} {\hat j}$ and ${\vec M}_2 = \frac{\sqrt 3}{2} {\hat i} -
\frac{3}{2} {\hat j}$ are spanning vectors for the reciprocal lattice. The 
operator $D_{\vn}$ can take the values $\pm 1$ independently for each $\vn$ 
and commutes with $H_F$, so that the states can be labeled by the values 
of $D_{\vn}$ on each bond; the ground state corresponds to 
$D_{\vn}=1$ on all bonds \cite{kitaev1,feng1,baskaran1,chen1} 
irrespective of the sign of $J_3$ due to a special symmetry of the 
model \cite{kitaev1}. Since $D_{\vn}$ is a constant of motion, the dynamics 
of the model starting from the ground state never takes the system outside 
the manifold of states with $D_{\vn}=1$. 

For $D_{\vn}=1$, Eq. (\ref{ham2}) can be diagonalized as $H_F = \sum_{\vk}
\psi_{\vk}^\da H_{\vk} \psi_{\vk}$, where $\psi_{\vk}^\da =(a_{\vk}^\da, ~
b_{\vk}^\da)$ are Fourier transforms of $a_{\vn}$ and $b_{\vn}$, the sum 
over $\vk$ extends over half the Brillouin zone (BZ) of the triangular 
lattice formed by the vectors $\vn$, and $H_{\vk}$ can be expressed
in terms of the Pauli matrices $\si^i$ (where $\si^3$ is diagonal) as
$H_{\vk} = 2 [J_1 \sin ({\vk} \cdot {\vec M}_1)-J_2 \sin
({\vk} \cdot {\vec M}_2)] \si^1 +2 [J_3 + J_1 \cos ({\vk} \cdot {\vec M}_1) +
J_2 \cos ({\vk} \cdot {\vec M}_2)] \si^2$. The spectrum
consists of two bands with energies $E_{\vk}^\pm = \pm E_{\vk}$, where
\bea E_{\vk} &=& 2 [ \{J_1 \sin ({\vk} \cdot {\vec M}_1) - J_2 \sin ({\vk}
\cdot {\vec M}_2) \}^2 \non \\
& & + \{ J_3 + J_1 \cos ({\vk} \cdot {\vec M}_1) + J_2 \cos ({\vk}
\cdot {\vec M}_2) \}^2 ]^{1/2} \label{hk1} \eea
For $|J_1-J_2|\le J_3 \le J_1+J_2$, the bands touch each other, and the
energy gap $\Delta_{\vk} = E_{\vk}^+ - E_{\vk}^-$ vanishes for special values
of $\vk$ leading to a gapless phase \cite{kitaev1,feng1,chen1,lee1}.

\begin{figure}
\rotatebox{0}{\includegraphics*[width=\linewidth]{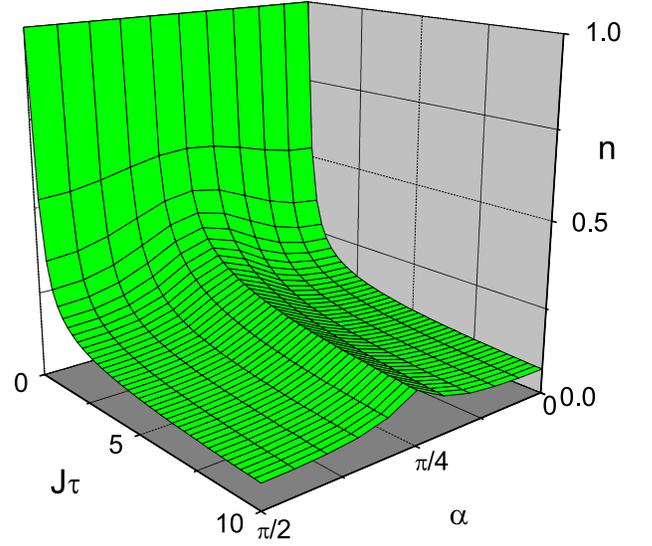}}
\caption{Plot of defect density $n$ versus $J \tau$ and $\al = \tan^{-1} 
(J_2/J_1)$. The density of defects is maximum at $J_1=J_2$.} \label{fig2} 
\end{figure}

We will now quench $J_3(t) =J t/\tau$ from $-\infty$ to $\infty$ at a fixed
rate $1/\tau$, keeping $J$, $J_1$ and $J_2$ fixed at some positive values.
The ground states of $H_F$ corresponding to $J_3 \to -\infty(\infty)$,
schematically shown in Fig. \ref{fig1}, are gapped and have $\si_{j,l}^z
\si_{j,l+1}^z = 1(-1)$ for all lattice sites $(j,l)$ of type $b$. To study
the time evolution of the system, we note that after an unitary
transformation $U= \exp(-i \si^1 \pi/4)$, we obtain $H_F
= \sum_{\vk} \psi_{\vk}^{'\da} H'_{\vk} \psi'_{\vk}$, where
$H'_{\vk} = UH_{\vk} U^\da$ is given by $ H'_{\vk} = 2 [J_1 \sin
({\vk} \cdot {\vec M}_1) - J_2 \sin ({\vk} \cdot {\vec M}_2)] \si^1
+ 2 [J_3(t) +J_1 \cos ({\vk} \cdot {\vec M}_1) + J_2 \cos ({\vk}
\cdot {\vec M}_2)] \si^3$. Hence the off-diagonal elements of
$H'_{\vk}$ remain time independent, and the quench
dynamics reduces to a Landau-Zener problem for each ${\vk}$. The
defect density can then be computed following a standard prescription
\cite{lz1}: $n = (1/A) ~\int_{\vk} ~d^2 \vk ~p_{\vk}$, where
\beq p_{\vk} = \exp[- 2 \pi \tau \{ J_1 \sin ({\vk} \cdot {\vec M}_1)- J_2
\sin ({\vk} \cdot {\vec M}_2) \}^2 /J] \eeq
is the probability of defect production for the state
labeled by momentum $\vk$, and $A = 4\pi^2 /(3\sqrt{3})$ denotes the area
of half the BZ over which the integration is carried out. A plot of $n$ as a
function of the quench time $J \tau$ and an angle $\al$ is shown in
Fig. \ref{fig2}; here we have taken $J_{1[2]} = J \cos \al [\sin \al]$. We
note that the density of defects produced is maximum when $\al = \pi /4$
($J_1=J_2$). This occurs because the length of the
gapless line through which the system passes during the quench is maximum
for $J_1=J_2$. Hence the system remains in the non-adiabatic state for the
maximum time during the quench, leading to the maximum density of defects.
Note that unlike some other models \cite{ks1}, the defects here do not 
correspond to topological defects since the dynamics always keeps $D_{\vn}=1$ 
on all bonds.

For sufficiently slow quench $2 \pi J \tau \gg 1$, $p_{\vk}$ 
is exponentially small for all values of ${\vk}$ except 
near the line $J_1 ~\sin ({\vk} \cdot {\vec M}_1) ~=~ J_2 \sin ({\vk}
\cdot {\vec M}_2)$; the contribution to the momentum integral
in the expression for $n$ comes from this region.
Note that the line $p_{\vk}=1$ precisely corresponds to the zeros
of the energy gap $\Delta_{\vk}$ as $J_3$ is varied for fixed $J_1, J_2$.
By expanding $p_{\vk}$ about this line, we see that for a very
slow quench, the defect density scales as $n \sim 1/\sqrt{\tau}$. This
demonstrates that the scaling of $n$ with $\tau$ crucially depends on the 
dimensionality of the critical surface since for a quench which takes the 
system through a {\it critical point} instead of a {\it critical line}, the 
defect density of the Kitaev model, which has $d=2$ and $\nu = z = 1$ 
\cite{comment1}, is expected to scale as $1/\tau$ \cite{anatoly1}. This 
observation leads to the following general conclusion.

Consider a $d$-dimensional model with $\nu=z=1$ which is described
by a Hamiltonian $H_d = \sum_{\vk} \psi^\da_{\vk} \left( \si^3
\epsilon(\vk) t/\tau + \Delta(\vk) \sigma^+ + \Delta^{\ast}(\vk)
\sigma^- \right) \psi_{\vk}$, where $\sigma^\pm$ $= (\sigma^1 \pm i
\sigma^2)/2$. Suppose that a quench takes the system through a
critical surface of $d-m$ dimensions. The defect density for a
sufficiently slow quench is given by \cite{lz1} $n = (1/A_d)
\int_{\rm BZ} d^d k e^{-\pi \tau f(\vk)} \simeq (1/A_d) ~ \int_{\rm
BZ} d^d k ~ \exp [~- \pi \tau \sum_{\al, \be=1}^m g_{\al \be}
k_{\al} k_{\be}] \sim 1/\tau^{m/2}$, where $A_d$ is the area of the
$d$-dimensional BZ, $f (\vk)=|\Delta(\vk)|^2/|\epsilon(\vk)|$
vanishes on the $d-m$ dimensional critical surface, $\al, \be$
denote one of the $m$ directions orthogonal to that surface, and
$g_{\al \be} = [\partial^2 f(\vk)/\partial k_{\al} \partial k_{\be}]_{\vk 
\in {\rm critical ~surface}}$. Note that this result depends only on the 
property that $f(\vk)$ vanishes on a $d-m$ dimensional surface, and not on 
the precise form of $f(\vk)$. For general values of $\nu$ and $z$, we note 
that a Landau-Zener type of scaling argument yields $\Delta \sim 
1/\tau^{z\nu/(z\nu+1)}$, where $\Delta$ is the energy gap \cite{anatoly1}. 
When one crosses a $d-m$ dimensional critical surface during the quench, 
the available phase space $\Omega$ for defect production scales as 
$\Omega \sim k^m \sim \Delta^{m/z} \sim 1/\tau^{m \nu/(z\nu +1)}$;
this leads to $n \sim 1/\tau^{m \nu/(z\nu +1)}$. For a quench through a 
critical point where $m=d$, we retrieve the results of Ref. \cite{anatoly1}.

Next we study the correlations between the defects produced during
the quench. To this end, we define the operators $O_{\vcr} ~=~ i b_{\vn}
a_{\vn + \vcr}$. In the spin language, $O_{\vec 0} = \si_{j,l}^z
\si_{j,l+1}^z$. For $\vcr \ne {\vec 0}$, $O_{\vcr}$ can be written as a
product of spin operators going from a $b$ site at $\vn$ to an
$a$ site at $\vn + \vcr$: the product begins with a $\si^x$ or $\si^y$ at
$(j,l)$ and ends with a $\si^x$ or $\si^y$ at $(j',l')$ with a string
of $\si^z$'s in between, where the forms of the initial and final $\si$
matrices depend on the positions of $j+l$ and $j'+l'$. Note that
for $J_3 \to -\infty(\infty)$, where the $z$ component of each spin is
locked with that of its vertically nearest neighbor, $\la
\psi_{-\infty(\infty)}| O_{\vcr} |\psi_{-\infty(\infty)} \ra =\pm
\delta_{\vcr, {\vec 0}}$. For the Kitaev model, it is known that
the spin correlations between sites lying on different bonds vanish,
{\it e.g.}, $\la \si_{a,\vn}^z \si_{b,\vec{n}+\vec{r}}^z \ra =0$ for $\vcr
\ne {\vec 0}$ \cite{baskaran1}. Therefore $\la O_{\vcr} \ra$ are the only
non-vanishing two-point correlators of the model \cite{chen1}. A non-zero
value of $\la O_{\vcr} \ra$ for $\vcr \ne {\vec 0}$ in the final state 
provides a signature of the defects. In particular, a plot of $\la O_{\vcr}
\ra$ versus $\vcr$ gives an estimate of the correlations between the defects.
[Since $O_{\vcr}^2 = 1$, all the moments of $O_{\vcr}$ can be found trivially:
$\la O_{\vcr}^n \ra = \la O_{\vcr} \ra$ if $n$ is odd and $=1$ if $n$ is even.]

After the quench, the system, for each momentum ${\vec k}$, is described
by a combination of $\psi_{-\infty\,\vk}$ with probability $p_{\vk}$ and
$\psi_{\infty \,\vk}$ with probability $1-p_{\vk}$, where $\psi_{\pm
\infty\,\vk}$ are the eigenstates of $H_{\vk}$ for $J_3 \to \pm \infty$.
Hence $\la O_{\vcr} \ra$ can be computed in a straightforward manner:
\beq \la O_{\vcr} \ra ~=~ - ~\de_{\vcr,{\vec 0}} ~+~ \frac{2}{A} ~ \int ~
d^2 \vk ~p_{\vk} ~\cos (\vk \cdot \vcr), \label{int2} \eeq
where the integral runs over half the BZ with area $A$.

For large values of $\tau$, the dominant contribution comes from the
region near the line $J_1 ~\sin ({\vk} \cdot {\vec M}_1) ~=~ J_2
\sin ({\vk} \cdot {\vec M}_2)$ where $p_{\vk}=1$. Introducing
the variables $k_\parallel$ and $k_\perp$ which vary along and
perpendicular to this line (along the directions ${\hat
n}_\parallel$ and ${\hat n}_\perp$ respectively), we see
that the integrand in Eq. (\ref{int2}) takes the form $\exp
[-a(\vk_0)\tau k_\perp^2 \pm i (\vk_0 + k_\parallel {\hat n}_\parallel +
k_\perp {\hat n}_\perp) \cdot \vcr]$, where $a(\vk_0)$ is a number depending
on $\vk_0$. The evaluation of the integral over $k_\perp$ gives a factor of
$\exp \left[ -(\vcr \cdot {\hat n}_\perp)^2 /(4a\tau) \right]/\sqrt{a \tau}$.
We thus see that the magnitude of the defect correlations goes as 
$1/\sqrt{\tau}$, while the spatial extent of the correlations goes as
$\sqrt{\tau}$. This is confirmed by the following relation:
$\sum_{\vcr} \vcr^2 \la O_{\vcr} \ra = -2 (\nabla^2_{\vk} p_{\vk})_{\vk = 
{\vec 0}} = 24 \pi \tau (J_1^2 + J_2^2 + J_1 J_2)/J$.

\begin{figure}
\rotatebox{0}{\includegraphics*[width=\linewidth]{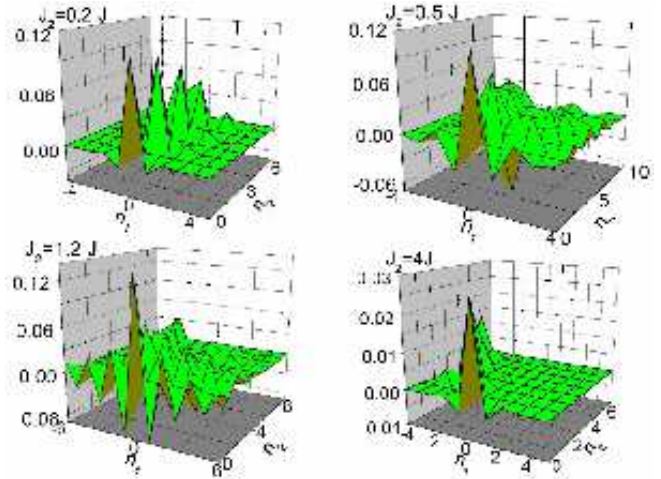}}
\caption{Plot of $\la O_{\vcr} \ra$, sans the $\de$-function peak at the 
origin, as a function of $\vcr$ for four values of $J_2/J$, for $J\tau=5$.}
\label{fig3} \end{figure}

To study the correlations between defects, we evaluate Eq. (\ref{int2}) 
numerically; the $\vcr$ dependence of $\la O_{\vcr} \ra$ is shown in Fig. 
\ref{fig3} for several values of $J_2/J_1$, where $J_1 =J$ and $J\tau=5$. 
Here we have omitted the $\de$-function peak at ${\vcr}=0$ in Eq. \ref{int2} 
so as to make the correlation at $\vcr \ne {\vec 0}$ clearly visible.
The plot of $\la O_{\vcr} \ra$ reflects the defect correlations. To
understand the variation of these correlations with the ratio
$J_2/J_1$, we note that for large $J\tau$, the maximum contribution
to $\la O_{\vcr} \ra$ comes from around the wave vectors ${\vec
k}_0$ for which $p({\vec k}_0)=1$. For $J_2 \gg (\ll) J_1$, this
implies $\sin[{\vec k}\cdot {\vec M}_2({\vec M}_1)] =0$ which yields
${\vec k}_0 \sim \sqrt{3} {\hat i} \pm {\hat j}$. The maximum
contribution to $\la O_{\vcr} \ra$ comes from $\cos({\vec k}_0 \cdot
\vcr) = 1$, {\it i.e.}, ${\vec k}_0 \cdot \vcr=0$. Thus for $J_2 \gg
(\ll) J_1$, $\la O_{\vcr} \ra$ is expected to be maximal along the
lines $y = -(+) \sqrt{3} x$, namely, $n_1 = - n_2$ ($n_1 = 0$) in
the $n_1-n_2$ plane. This expectation is confirmed in Fig.
\ref{fig3} where $\la O_{\vcr} \ra$ can be seen to be maximal along
$n_1 = - n_2$ ($n_1 = 0$) for $J_2=4(0.2)J_1$. Such a strong
anisotropy can be understood by noting that the
Kitaev model reduces to a one-dimensional model when $J_2 \gg (\ll)
J_1$. For intermediate values of $J_1/J_2$, a gradual evolution of
the defect correlations can be seen in Fig. \ref{fig3}. We note that if
the Kitaev model can be realized using ultracold atoms in an optical
lattice \cite{expt_real}, such an evolution of the defect correlations 
with $J_1/J_2$ can, in principle, be experimentally detected by spatial
noise correlation measurements as pointed out in Ref. \cite{altman1}.

\begin{figure}
\rotatebox{0}{\includegraphics*[width=\linewidth]{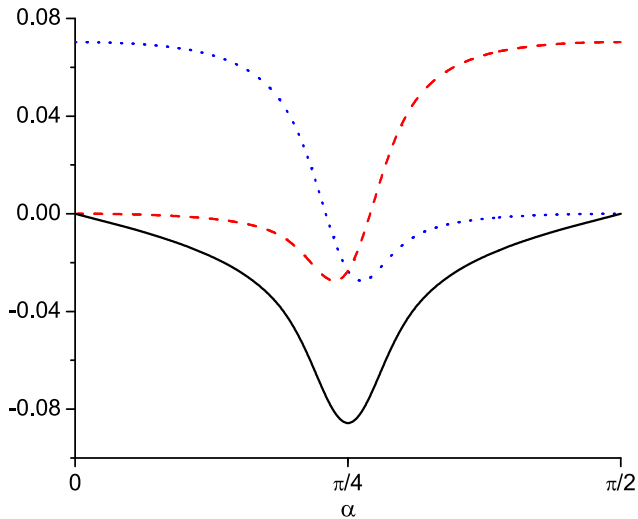}}
\caption{Plot of $\la O_{\vcr} \ra$ at the points $(1,0)$ on the $n_1$ axis 
(black solid line), $(0,2)$ on the $n_2$ axis (blue dotted line), and 
$(2,-2)$ along the $-45^{\circ}$ line in the $n_1-n_2$ plane (red dashed 
line) as a function of $\al = \tan^{-1} (J_2/J_1)$, for $J^2=1$ and 
$J\tau =5$.} \label{fig4} \end{figure}

We can obtain a different view of the spatial anisotropy of the defect
correlations by studying $\la O_{\vcr} \ra$
as a function of $\al = \tan^{-1} (J_2 /J_1)$. As $\al$ changes, the ratio
$J_2/J_1$ varies from $0$ to $\infty$ while fixing $J_1^2+J_2^2=J^2=1$. A
plot of $\la O_{\vcr} \ra$ at three representative points $(n_1,n_2)= (1,0)$
(on the $n_1$ axis), $(0,2)$ (on the $n_2$ axis), and $(2,-2)$ (along the
$-45^{\circ}$ line in the $n_1-n_2$ plane) as a function of $\al$, shown in
Fig. \ref{fig4}, reveals the nature of the defect correlations. We see that 
as $J_2/J_1 = \tan \al$ is varied from $0$ to $\infty$, the magnitude of
the correlation at the point $(1,0)$ on the $n_1$ axis increases till it 
reaches a maximum at $J_1=J_2$ ($\al=\pi/4$), and then decays to $0$ as 
$\al$ approaches $\pi/2$. For the points $(0,2)$ on the $n_2$ axis and 
$(2,-2)$ along the line with slope $-45^{\circ}$, the correlation becomes 
maximum when $J_2 \ll J_1$ ($\al =0$) and $J_2 \gg J_1$ ($\al=\pi/2$) 
respectively, as expected from Fig. \ref{fig3}. We conclude that the spatial 
anisotropy of the correlations between the defects $\la O_{\vcr} \ra$ 
depends crucially on the ratio $J_2/J_1$. 

To conclude, we have shown that the density of defects produced
during a quench of the Kitaev model through a critical line scales
with the quench time as $1/\sqrt{\tau}$, instead of the $1/\tau$
behavior expected for a quench through a critical point. We have provided
a general result for the defect density which reproduces the result of
Ref. \cite{anatoly1} as a special case. We have also discussed the variation
of the defect correlations with the model parameters and pointed out the 
possibility of detection of these variations in experiments. These
results significantly improve our general understanding of the scaling of
the density of defects and their correlations in 2D systems.

We thank A. Dutta and A. Polkovnikov for stimulating discussions.

\end{document}